# Fe-assisted epitaxial growth of 4-inch single-crystal transition-metal dichalcogenides on *c*-plane sapphire without miscut angle

Hui Li[1,#], Junbo Yang[1,#], Xiaohui Li[1,#], Mo Cheng[1], Wang Feng[1], Ruofan Du[1], Yuzhu Wang[1], Luying Song[1], Xia Wen[1], Lei Liao[2], Yanfeng Zhang[3], Jianping Shi[1]*, Jun He[4]*

[1]The Institute for Advanced Studies, Wuhan University, Wuhan, China

[2]College of Semiconductors, School of Physics and Electronics, Hunan University, Changsha, China

[3]School of Materials Science and Engineering, Peking University, Beijing, China

[4]Key Laboratory of Artificial Micro- and Nano-structures of Ministry of Education, School of Physics and Technology, Wuhan University, Wuhan, China

*Address correspondences: jianpingshi@whu.edu.cn, He-jun@whu.edu.cn

#These authors contributed equally to this work.

**Epitaxial growth and controllable doping of wafer-scale single-crystal transition-metal dichalcogenides (TMDCs) are two central tasks for extending Moore's law beyond silicon. However, despite considerable efforts, addressing such crucial issues simultaneously under two-dimensional (2D) confinement is yet to be realized. Here we design an ingenious epitaxial strategy to synthesize record-breaking 4-inch single-crystal Fe-doped TMDCs monolayers on industry-compatible *c*-plane sapphire without miscut angle. In-depth characterizations and theoretical calculations reveal that the introduction of Fe significantly decreases the formation energy of parallel steps on sapphire surfaces and contributes to the edge-nucleation of unidirectional TMDCs domains (>99%). The ultrahigh electron mobility (~86 cm$^2$ V$^{-1}$ s$^{-1}$) and remarkable on/off current ratio (~10$^8$) are discovered on 4-inch single-crystal Fe-MoS$_2$ monolayers due to the ultralow contact resistance (~489 Ω μm) and perfect Ohmic contact with metal electrodes. This work represents a substantial leap in terms of bridging the synthesis and doping of wafer-scale single-crystal 2D semiconductors without the need for substrate miscut, which should promote the further device downscaling and extension of Moore's law.**

Two-dimensional (2D) transition-metal dichalcogenides (TMDCs) reveal an unprecedented potential to consistently drive advanced performance of integrated circuit and extend Moore's law beyond the 2-nm node, in view of their atomically thin thicknesses, flexible band structures, and silicon-compatible features[1–8]. Nevertheless, many fundamental challenges should be resolved before 2D TMDCs meet the industry criteria for practical application in electronic device. For example, the large-area growth of single crystals to decrease defective grain boundaries[9,10], the effective doping of 2D TMDCs to modulate band structures[11,12], and the optimization design of interfaces to reduce contact resistances[13–17]. Therein, controllable synthesis and effective doping of wafer-scale 2D TMDCs single crystals are two central tasks for extending Moore's law beyond silicon. Although the inch sized single-crystal graphene[18–20] and hexagonal boron nitride ($h$-BN)[21–24] have been synthesized on metal substrates, such strategies are incompatible with the growth of 2D TMDCs on insulating substrates due to the different kinetics.

Noticeably, 1-inch single-crystal 2H-MoTe$_2$ films have synthesized on SiO$_2$/Si by means of phase transition and recrystallization[25]. In addition, the step-edge-guided growth of single-crystal 2D TMDCs has also been proposed on sapphire substrate. For example, 2-inch single-crystal MoS$_2$ and WS$_2$ monolayers have been synthesized on custom-designed $c/a$ and $a$-plane sapphire, where the nucleation of MoS$_2$ and WS$_2$ are along with the step edges of substrates[26,27]. By engineering the atomic terrace height on $c/a$ sapphire, centimeter-sized uniform bilayer MoS$_2$ films have also been obtained[28]. However, the direction and height of step edge on sapphire are dependent on the miscut direction, and maintaining a fixed cutting angle with such a high accuracy over inch scale is difficult. Another point of concern, controlled doping of 2D semiconductors is crucial to modulate band structures and induce novel physical phenomena. However, excessive doping should introduce the deep impurity level and degrade the device's performance. Interestingly, by designing a remote modulation doping strategy, the 2D confined charge transport and the suppression of impurity scattering were achieved in WSe$_2$/MoS$_2$ heterostructure[29]. The incorporation of oxygen in monolayer MoS$_2$ could be used to passivate sulfur vacancy and suppress the formation of donor state[30]. Nevertheless, to the best of our knowledge, the effective doping of wafer-scale single-crystal 2D TMDCs is yet to be realized.

Herein, we design a Fe-assisted chemical vapor deposition (CVD) method to synthesize 4-inch single-crystal Fe-doped TMDCs monolayers on industry-compatible $c$-plane sapphire without miscut angle. The superiorities of such a strategy can be summarized as follows: (1) the introduction of Fe contributes to the formation of parallel steps on sapphire surfaces and the edge-nucleation of unidirectional monolayer TMDCs domains; (2) the employment of $c$-plane sapphire without miscut angle significantly reduces the preparation cost and enables the industry application;

(3) the controlled doping of Fe modulates the band structures of TMDCs and delivers excellent device performances. This work provides an innovative strategy for bridging the synthesis and doping of wafer-scale single-crystal 2D semiconductors, which enables further device downscaling and extension of Moore's law.

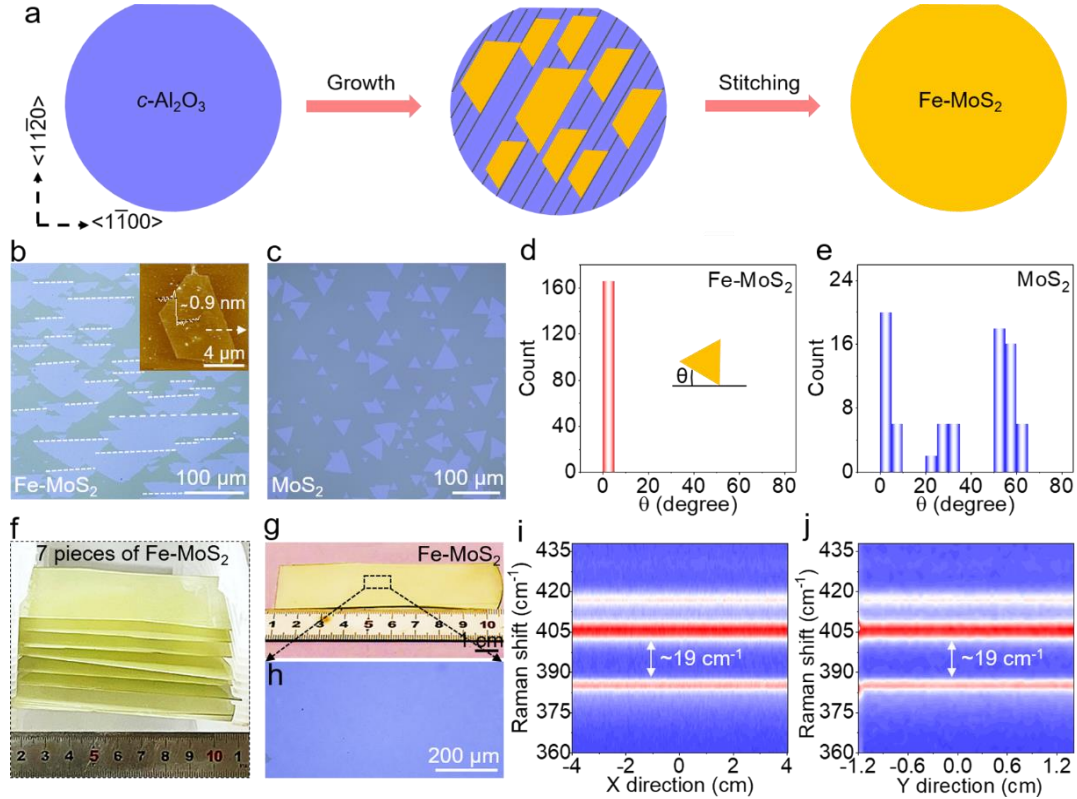

*Fig. 1 | Epitaxial growth of 4-inch single-crystal Fe-MoS$_2$ monolayers on c-plane sapphire without miscut angle. **a**, Schematic diagram of the growth process of 4-inch single-crystal MoS$_2$ monolayers on industry-compatible c-plane sapphire. **b,c**, OM images of Fe-MoS$_2$ and MoS$_2$ on c-plane sapphire, showing the unidirectionally aligned feature of Fe-MoS$_2$. Inset of (**b**): AFM image and corresponding height profile of a single Fe-MoS$_2$ domain, revealing its monolayer feature. **d,e**, Statistical analyses of domain orientations of Fe-MoS$_2$ and MoS$_2$ on c-plane sapphire. **f**, Photograph of 7 pieces of single-crystal Fe-MoS$_2$ monolayer/c-plane sapphire. **g**, Photography a 4-inch single Fe-MoS$_2$ monolayer on c-plane sapphire. **h**, OM image captured from the marked region in (**g**). **i,j**, Color-code images of Raman line scan mapping along the horizontal and longitudinal direction, respectively, showing the ultrahigh thickness uniformity of monolayer Fe-MoS$_2$ film.*

## Results

**Epitaxial growth of 4-inch single-crystal Fe-MoS$_2$ on *c*-plane sapphire.** Sapphire is a frequently used growth substrate for synthesizing 2D materials[31,32] and the formation of parallel steps on its surface contributes to the evolution of unidirectionally aligned domains[26–28]. Nevertheless, the time-consuming high temperature annealing process is essential for producing such parallel steps[33]. Notably, an ingenious Fe-assisted CVD method is developed for growing unidirectionally aligned monolayer TMDCs domains and single-crystal films on 4-inch commercial *c*- plane sapphire, and the parallel steps are formed rapidly at relatively low growth temperature (Fig. 1a and Supplementary Fig. 1). Figure 1b and Supplementary Fig. 2 show the typical optical microscopy (OM) images of Fe-MoS$_2$ domains with nearly 100% unidirectional alignment on $c$-plane sapphire, different from the pristine MoS$_2$ with random orientations (Fig. 1c). This is further corroborated by the statistical analyses of domain orientations in Fig. 1d,e. Such results indicate that the introduction of Fe contributes to the formation of parallel steps on sapphire surfaces and subsequent synthesis of unidirectional alignment domains. The atomic force microscopy (AFM) image and corresponding height profile in Fig. 1b reveal that the thickness of Fe-MoS$_2$ is ~0.9 nm, corresponds to the monolayer. To further determine Fe doping in monolayer MoS$_2$, X-ray photoelectron spectroscopy (XPS) measurements were then performed on as-grown samples (Supplementary Fig. 3). The binding energy of ~712.7 eV is attributed to Fe$^{4+}$, consistent with the XPS result of Fe-doped SnSe$_2$[34], and the Fe doping concentration is determined to be ~4%. In addition, comparative Raman and photoluminescence (PL) investigations were also carried out on monolayer Fe-MoS$_2$ and pristine monolayer MoS$_2$. A red-shift of A$_{1g}$ mode is obviously observed in monolayer Fe-MoS$_2$ (~401.4 cm$^{-1}$) comparing with that of the pristine monolayer MoS$_2$ (~404.4 cm$^{-1}$), indicative of the n-type doping effect of Fe on monolayer MoS$_2$ (Supplementary Fig. 4a), in line with the result of CVD-synthesized monolayer Fe-MoS$_2$ on SiO$_2$/Si[35]. Meanwhile, the reduced PL intensity is obtained in monolayer Fe-MoS$_2$ due to the increased electron and exciton recombination (Supplementary Fig. 4b).

Further increasing the growth time from ~5 to ~60 minutes, 4-inch single-crystal monolayer Fe-MoS$_2$ films are successfully synthesized on $c$-plane sapphire, as shown in Fig. 1f–h and Supplementary Fig. 5. Figure 1f manifests the digital photography of 7 pieces of 4-inch monolayer Fe-MoS$_2$ films, showing an identical color reflection as a symbol of macroscopic homogeneity. Furthermore, by using the Raman line scanning, the wafer-scale thickness uniformity of monolayer Fe-MoS$_2$ film is investigated (Fig. 1i,j), and nearly the same Raman peak positions along the entire wafer diameter suggests the ultrahigh uniformity. Briefly, by using a Fe-assisted CVD method, 4-inch single-crystal Fe-MoS$_2$ monolayers have been synthesized on industry-compatible $c$-plane sapphire without miscut angle, which provides an ideal platform for constructing high-performance electronic devices.

**Multiscale determining the single-crystal feature of monolayer Fe-MoS$_2$.** To further identify the single-crystal feature of CVD-derived monolayer Fe-MoS$_2$, multiscale characterizations were performed on as-grown and transferred samples. The low-magnification transmission electron microscopy (TEM) image captured from the merged region of two unidirectional alignment Fe-MoS$_2$ domains is presented in Fig. 2a, the well-preserved morphology and transparent property suggest the high crystalline quality and ultrathin feature. The corresponding selected-area electron diffraction (SAED) pattern shows only one set of hexagonally arranged diffraction spots, suggestive of the perfect coalescence of such two domains (Fig. 2b). Furthermore, a series of SAED patterns obtained from different regions of a monolayer Fe-MoS$_2$ film reveal nearly identical lattice orientation, highly indicative of its single-crystal feature (Supplementary Fig. 6). The atomic-resolution high-angle annular dark-field scanning transmission electron microscopy (HAADF-STEM) images clearly present the atomic evidence of seamless stitching between the adjacent Fe-MoS$_2$ domains (Fig. 2c–g). In addition, the energy dispersive X-ray spectroscopy (EDS) measurements were also performed to identify the chemical constitutions and their distributions (Supplementary Fig. 7). The uniform color contrast within the domain reveals the high crystalline quality of monolayer Fe-MoS$_2$.

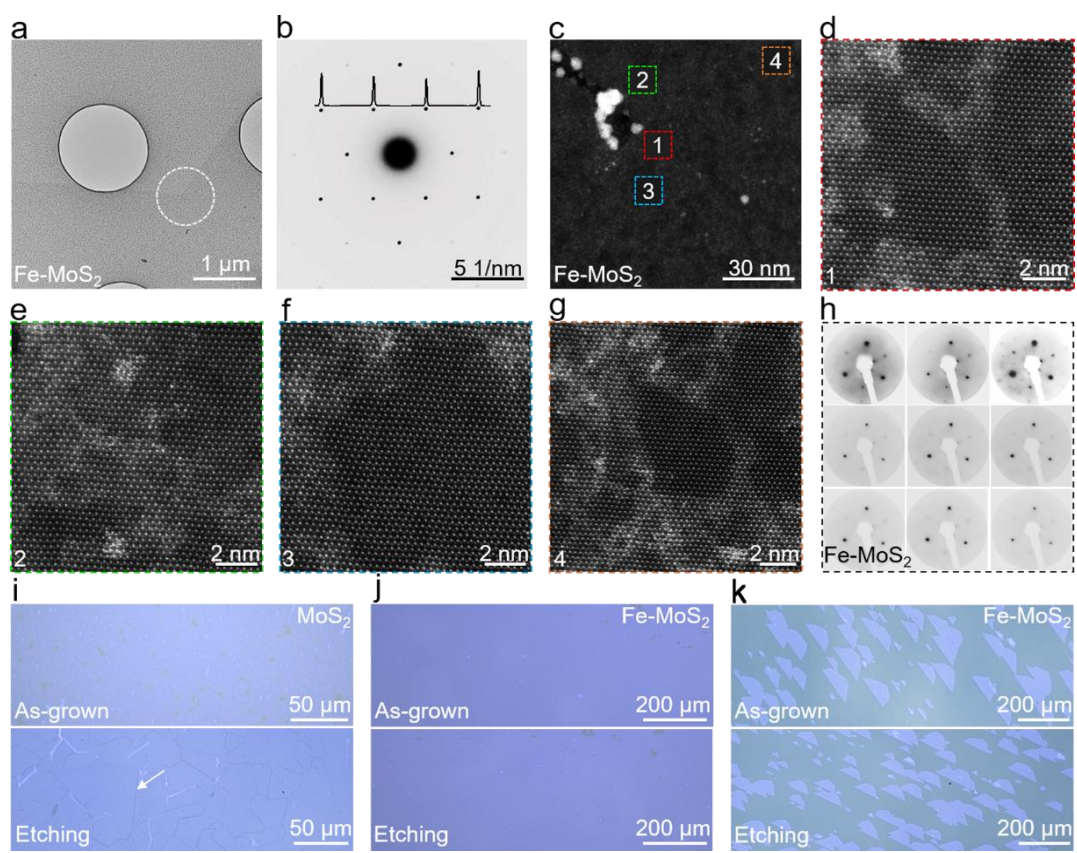

*Fig. 2 | Multiscale characterizations of the single-crystal feature of monolayer Fe-MoS$_2$. a,* Low-magnification TEM image collected from the merged area of two unidirectional alignment Fe-MoS$_2$ domains. *b,* Corresponding SAED pattern captured from the white circle in *(a)*. *c,* Low-magnification STEM image captured from the merged region of two unidirectional alignment Fe-MoS$_2$ domains. *d–g,* Atomic-resolution HAADF-STEM images obtained from the marked areas in *(c)*. *h,* LEED patterns of monolayer Fe-MoS$_2$ film scanning across a ~2 cm$^2$ sample area. The electron energy is 230 eV and the spot size is ~1 mm. *i,* OM images of as-grown polycrystalline monolayer MoS$_2$ films on sapphire before and after O$_2$ etching at 400 °C. *j,k.* OM images of as-grown monolayer Fe-MoS$_2$ films and unidirectionally aligned domains on sapphire before and after O$_2$ etching at 400 °C.

Figure 2h reveals the low-energy electron diffraction (LEED) patterns measured at nine random locations across the ~2 cm$^2$ monolayer Fe-MoS$_2$ film. The homogeneous orientations and three bright spots at certain voltage unambiguously indicate the single-crystal feature of CVD-synthesized monolayer Fe-MoS$_2$, as well as its well-defined epitaxial relationship with sapphire substrate. To further verify this conclusion at macroscale, the O$_2$ etching experiments were then performed[36–38], where the grain boundaries were obviously observed for CVD-synthesized polycrystalline MoS$_2$ films (Fig. 2i). However, for unidirectional aligned Fe-MoS$_2$ domains and films, almost no contrast variation is presented, highly suggestive of the single-crystal property (Fig. 2j,k). Overall, the wafer-scale single-crystal monolayer Fe-MoS$_2$ film can be achieved by the seamless stitching of unidirectionally aligned domains, and has been verified by the multiscale characterization results.

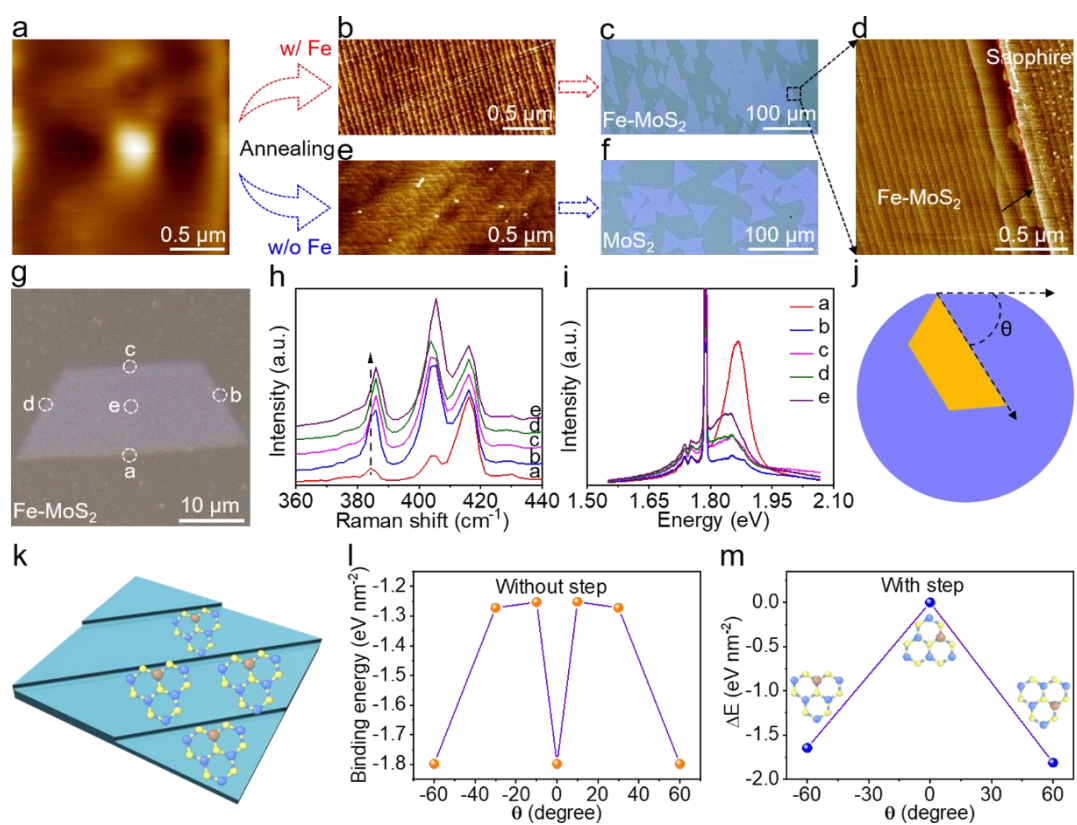

*Fig. 3 | The epitaxial growth mechanism of unidirectionally aligned Fe-MoS$_2$ on c-plane sapphire. **a**, AFM image of an initial c-plane sapphire. **b,e**, AFM images of c-plane sapphire annealing at ~900 °C with and without the presence of FeCl$_2$. **c,f**, OM images of monolayer Fe-MoS$_2$ and pristine monolayer MoS$_2$ on c-plane sapphire. **d**, AFM image captured from the long domain edge of monolayer Fe-MoS$_2$. Fe-MoS$_2$ domain is nucleated at the parallel step edge of sapphire. **g**, OM image of as-grown monolayer Fe-MoS$_2$ on c-plane sapphire. **h,i**, Corresponding Raman and PL spectra captured from different regions of Fe-MoS$_2$, suggesting the adsorption of Fe particles at the step edges of sapphire. **j**, Schematic diagram of the configuration of Fe-MoS$_2$ on c-plane sapphire, where θ is the angle between the long domain edge of Fe-MoS$_2$ and the <1–100> direction of c-plane sapphire. **k**, Schematic diagram of Fe-MoS$_2$ islands on c-plane sapphire in the early stage of growth. The driving force of edge coupling between Fe-MoS$_2$ and c-*

*plane sapphire step results in the unidirectional alignment of islands. **l,m**, DFT calculated the binding energies of Fe-MoS$_2$ island with different rotation angles on c-plane sapphire.*

**The growth mechanism of unidirectionally aligned Fe-MoS$_2$.** The parallel step on sapphire surface plays a pivotal role for reducing the symmetry and synthesizing unidirectionally aligned TMDCs[26–28,39,40]. To understand the growth mechanism of unidirectionally aligned Fe-MoS$_2$ domains on *c*-plane sapphire, in-depth experimental characterizations and theoretical calculations were thus performed. The AFM image of an initial *c*-plane sapphire is shown in Fig. 3a, in which no step is observed. However, after annealing at ~900 °C with the presence of FeCl$_2$, perfect parallel steps are clearly obtained (Fig. 3b and Supplementary Fig. 8a–c), and which determine the subsequent growth of unidirectionally aligned Fe-MoS$_2$ domains, as presented in Fig. 3c. Interestingly, monolayer Fe-MoS$_2$ domains are found to be nucleated at the step edges and then growth along with the <10–10> direction of sapphire (Fig. 3d). In addition, a bright line (indicated by the black arrow) is clearly observed at the step edge, possibly originates from the adsorbed Fe particles. However, for the *c*-plane sapphire annealing at ~900 °C without the presence of FeCl$_2$, distorted steps are appearing (Fig. 3e and Supplementary Fig. 8d–f), and which induce the synthesis of MoS$_2$ domains with random orientations, as shown in Fig. 3f. Such results indicate that the introduction of Fe significantly decreases the formation energy of parallel steps on *c*-plane sapphire surfaces, which is confirmed by the relatively low annealing temperature (~900 °C for 5 minutes) comparing with those of *c/a* (~1000 °C for 4 hours)[26] and *a*-plane (~1140 °C for 5 hours)[27] sapphire. This is crucial for the epitaxial growth of unidirectionally aligned TMDCs domains and single-crystal films.

Raman and PL characterizations were then performed to identify the adsorbed Fe particles on the step edges of *c*-plane sapphire surfaces (Fig. 3g–i). Notably, a red-shift of $E_{2g}^1$ mode is obviously observed for the Raman spectrum captured from the long domain edge of monolayer Fe-MoS$_2$ (~384.2 cm$^{-1}$), which along with the step edge of sapphire, comparing with the other domain edges and base-plane (~386.0 cm$^{-1}$), as shown in Fig. 3h. This phenomenon can be understood by the strong stress between adsorbed Fe particles and domain edges of monolayer MoS$_2$[41]. Furthermore, in view of the plasmon enhanced effect from the adsorbed Fe particles and the strong interaction with step edges, the increased PL intensity and blue-shift peak are observed on the long domain edge of monolayer Fe-MoS$_2$ (Fig. 3i)[42]. Such results suggest that Fe particles preferentially adsorb at the step edges of sapphire surfaces and contribute to the nucleation of unidirectionally aligned MoS$_2$. Density functional theory (DFT) calculations were also carried out to further explore the growth mechanism of monolayer Fe-MoS$_2$ on *c*-plane sapphire. The binding energies of Fe-MoS$_2$ island with different rotation angles on *c*-plane sapphire are achieved (Fig. 3j–m). Notably, two antiparallel alignments of Fe-MoS$_2$ islands are evolved on *c*-plane sapphire without the assistance of parallel steps, which should induce the appearance of twin boundaries (Fig. 3l). Nevertheless, the introduction of parallel steps on *c*-plane sapphire surfaces breaks the degeneracy of such two antiparallel alignments and contributes to the synthesis of single crystals (Fig. 3m). Furthermore, the adsorbed Fe particles at the step edges are also beneficial for the nucleation of unidirectionally aligned MoS$_2$ domains, as described in Fig. 3k.

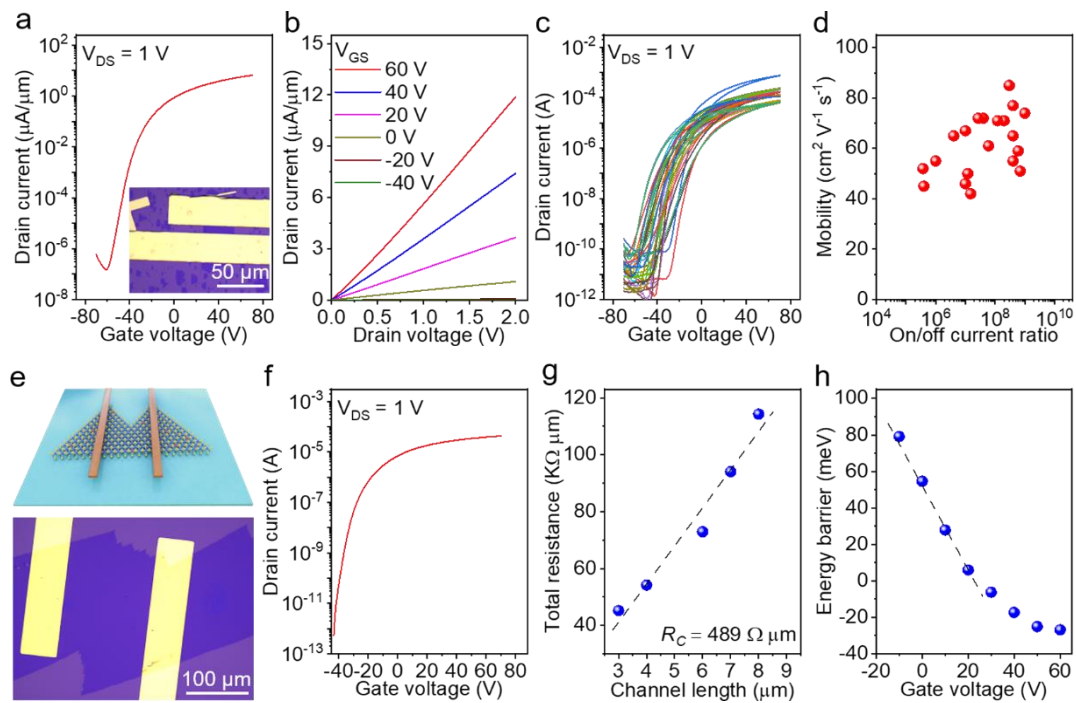

**Fig. 4 | *The device performance of monolayer Fe-MoS$_2$*. a**, Transfer characteristic curve of a monolayer Fe-MoS$_2$ back-gated FET. Inset: OM image of the device. **b**, Corresponding output characteristic curves. **c**, Transfer characteristic curves of 20 monolayer Fe-MoS$_2$ FETs measured across a 4-inch wafer. **d**, Statistical distribution of electron mobility and on/off current ratio. **e**, Schematic diagram and OM image of monolayer Fe-MoS$_2$ FET across the merged area of two unidirectionally aligned domains. **f**, Corresponding transfer characteristic curve. **g**, Contact resistance extraction of monolayer Fe-MoS$_2$ using the transfer-length method. **h**, Energy barrier of monolayer Fe-MoS$_2$.

**The device performances of Fe-MoS$_2$.** Monolayer Fe-MoS$_2$ field-effect transistors (FETs) were fabricated by using Cr/Au as the electrodes to evaluate the device performances (inset of Fig. 4a). The n-channel conduction is convinced by the transfer characteristic curve in Fig. 4a, consistent with the Raman results. More interestingly, the ultrahigh electron mobility (~86 cm$^2$ V$^{-1}$ s$^{-1}$) and excellent on/off current ratio (~10$^8$) are obtained, much higher than those of monolayer MoS$_2$ nanosheets[15,29,30] and polycrystalline films[43–45]. The comparison of electron mobility of monolayer Fe-MoS$_2$ with the other monolayers is shown in Supplementary Table 1. Meanwhile, the typical linear output characteristic curves of monolayer Fe-MoS$_2$ FET at room temperature suggest the perfect Ohmic contact between the channels and Cr/Au electrodes (Fig. 4b). Figure 4c shows the transfer curves of 20 monolayer Fe-MoS$_2$ FETs, the high device yield and uniform device performance are confirmed by the relatively small current variation. The statistical analysis of electron mobility and on/off current ratio is revealed in Fig. 4d, the average and best mobilities are obtained to be ~62 and ~86 cm$^2$ V$^{-1}$ s$^{-1}$, respectively. The relatively small variation of electron mobility reconfirms the ultrahigh crystalline quality and uniformity of 4-inch single-crystal Fe-MoS$_2$ monolayer.

To further confirm the seamless stitching of unidirectionally aligned monolayer Fe-MoS$_2$, a large-channel (~100 μm) FET was thus constructed by depositing Cr/Au electrodes on different domains, as described in Fig. 4e. Interestingly, the relatively high electron mobility (~48 cm$^2$ V$^{-1}$ s$^{-1}$) and on/off current ratio (~10$^8$) are still obtained, highly suggestive of its single-crystal feature (Fig. 4f and Supplementary Fig. 9a). The contact resistance is extracted by using the transfer-length method (TLM), as shown in Fig. 4g and Supplementary Fig. 9b. Notably, the contact resistance of monolayer Fe-MoS$_2$ is calculated to be ~489 Ω μm, much smaller than that of the monolayer MoS$_2$ (Supplementary Table 2), convincing the outstanding contact with Cr/Au electrodes. The internal mechanism of enhanced device performance in monolayer Fe-MoS$_2$ is thus clarified by executing the temperature-dependent electrical measurements (Supplementary Fig. 10a,b). With decreasing the temperature from ~300 to ~40 K, the increased current density and electron mobility are observed due to the reduced thermionic emission current between monolayer Fe-MoS$_2$ and electrodes. Meanwhile, the negative correlation between electron mobility and temperature suggests the suppressed ionized impurity scattering. The energy barrier of monolayer Fe-MoS$_2$ is also extracted from the Arrhenius plots (Fig. 4h and Supplementary Fig. 10c) and the zero Schottky barrier height indicates the perfect contact between monolayer Fe-MoS$_2$ and electrodes.

**Epitaxial growth of 4-inch single-crystal Fe-WS$_2$ monolayers.** Notably, the Fe-assisted epitaxial strategy is universal and can be used for synthesis of other wafer-scale TMDCs single crystals. A photograph of 4-inch Fe-WS$_2$ film is manifested in Fig. 5a, and the identical color contrast suggests its macroscopic thickness homogeneity. Notably, unidirectionally aligned monolayer Fe-WS$_2$ domains are evolved on *c*-plane sapphire (Fig. 5b and Supplementary Fig. 11), consistent with the result of monolayer Fe-MoS$_2$, reconfirming the prominent role of Fe in the epitaxial growth of single crystals. After increasing the growth time to 30 minutes, the 4-inch monolayer Fe-WS$_2$ film is obtained (Fig. 5c) and the thickness uniformity is convinced by nearly the same Raman peak positions (Fig. 5d).

LEED patterns in Fig. 5e clearly verify the single-crystal feature of monolayer Fe-WS$_2$ film due to the homogeneous orientations. In addition, the SAED and atomic-resolution HAADF-STEM images provide the microcosmic proof of the seamless stitching between neighboring Fe-WS$_2$ domains, which further clarifies the single-crystal nature of Fe-WS$_2$ (Fig. 5f–j and Supplementary Fig. 12). The Fe doping concentration is determined to be ~5%, according to the XPS and EDS results (Supplementary Figs. 13 and 14). Furthermore, the n-type doping of Fe and the robust atomic structure is also confirmed by Raman and PL characterizations (Supplementary Fig. 15).

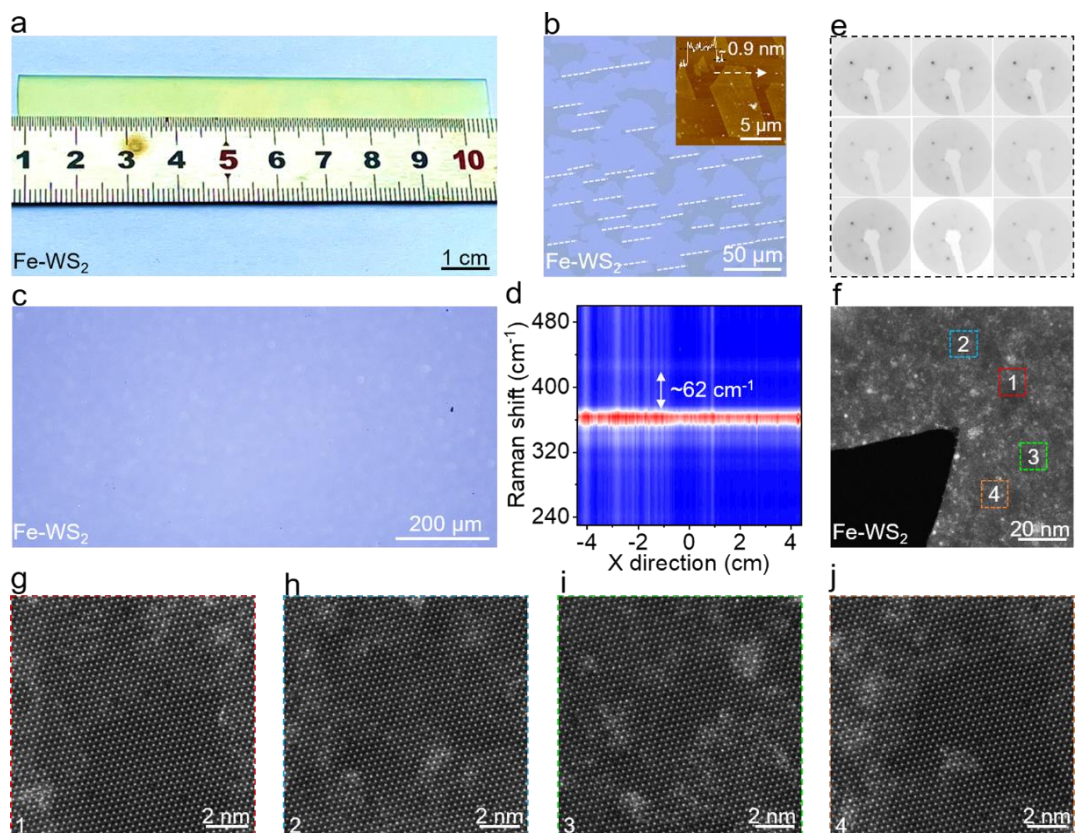

*Fig. 5 | Epitaxial growth of 4-inch single-crystal Fe-WS$_2$ monolayers on c-plane sapphire without miscut angle. a*, Photograph of a 4-inch single-crystal Fe-WS$_2$ on c-plane sapphire. *b*, OM image of Fe-WS$_2$ on c-plane sapphire. Inset: AFM image and corresponding height profile of a single Fe-WS$_2$ domain, revealing its monolayer feature. *c*, OM image of monolayer Fe-WS$_2$ film. *d*, Color-code image of Raman line scan mapping along the horizontal direction, showing the ultrahigh thickness uniformity of monolayer Fe-WS$_2$. *e*, LEED patterns of monolayer Fe-WS$_2$ film scanning across a ~2 cm$^2$ sample area. The electron energy is 230 eV and the spot size is ~1 mm. *f*, Low-magnification STEM image of the merged region of two unidirectional alignment Fe-WS$_2$ domains. *g–j*, Atomic-resolution HAADF-STEM images obtained from the marked areas in (*f*).

**Discussion**

In summary, we have developed a general Fe-assisted epitaxial strategy for synthesizing 4-inch single-crystal Fe-doped TMDCs monolayers on industry-compatible *c*-plane sapphire without miscut angle. The introduction of Fe reduces the formation temperature of parallel steps on sapphire surfaces and contributes to the edge-nucleation of unidirectional TMDCs domains. The distinguished device performances are demonstrated at wafer-scale, featured with ultrahigh electron mobility and excellent on/off current ratio. The controlled synthesis and effective doping of wafer-scale 2D TMDCs single crystals on low-cost *c*-plane sapphire provide the cornerstone for the device application toward the end-of-roadmap transistor.

**Methods**

**Epitaxial growth of 4-inch single-crystal Fe-doped TMDCs monolayers.** The 4-inch single-crystal TMDCs monolayers were synthesized in a dual-heating-zone furnace. Before executing the CVD growth, the *c*-plane sapphire substrates were cleaned with water, ethanol, and acetone, respectively. The $FeCl_2$ (~99.99%, Alfa Aesar) and $MoO_3$ (~99.99%, Alfa Aesar) were used as the precursors and placed downstream of S powders (~99.99%, Alfa Aesar). The heating rate was set as ~13 °C min$^{-1}$ and kept at ~900 °C for ~5 minutes to grow unidirectionally aligned Fe-$MoS_2$ domains. After increasing the growth time to ~60 minutes, 4-inch monolayer Fe-$MoS_2$ films were obtained. When the temperature reached to ~400 °C, the S powders were heated. After completing the CVD growth process, the furnace cover was opened and cooled down to room temperature. Notably, for CVD synthesis of monolayer Fe-$WS_2$, the growth temperature was set as ~950 °C. To trace the formation of parallel steps on sapphire surfaces, $FeCl_2$ was placed upstream of *c*-plane sapphire and annealed under Ar atmosphere at ~900 °C for ~5 minutes. The disordered steps were obtained on *c*-plane sapphire under the same condition without $FeCl_2$.

**Transfer and characterization of 4-inch single-crystal Fe-doped TMDCs monolayers.** CVD-synthesized monolayer Fe-$MoS_2$ and Fe-$WS_2$ were transferred by using the polymethyl methacrylate (PMMA) assistant method[46]. The morphology, domain size, thickness, optical property, and crystalline quality of Fe-$MoS_2$ and Fe-$WS_2$ were characterized by OM (Olympus BX53M), AFM (Dimension Icon, Bruker), XPS (ESCALAB250Xi, Mg Kα as the excitation source), LEED (BDL600-3GR), Raman spectroscopy (Renishaw with the excitation light of ~532 nm), and TEM (JEOL JEM-F200 and JEM-NEOARM with the acceleration voltage of ~200 kV). The atomic resolution HAADF-STEM and EDS results were obtained from a spherical-aberration-corrected STEM JEOL ARM200F with an acceleration voltage of ~80 kV.

**Device fabrication and measurement.** The ultraviolet lithography was used to pattern the source and drain. The electron beam evaporation system was employed for depositing Cr/Au electrodes with the thickness of 5/70 nm. The electrical transport measurements were performed under the vacuum (<1.3 mTorr) and dark conditions by using a semiconductor characterization system (Keithley 4200-SCS).

**Calculation of the electron mobility.** The electron mobility of monolayer Fe-MoS$_2$ was obtained based on the transfer curves according to the following expression:

$$\mu = \frac{L(dI_{DS}/dV_{GS})}{WC_{OX}V_{DS}}$$

Where $L$ and $W$ were the length and width of channel, and $C_{OX}$ was the gate capacitance of SiO$_2$, which could be calculated by $C_{OX} = \varepsilon_0\varepsilon_r/d$ ($\varepsilon_0$ was the vacuum permittivity, $\varepsilon_r$ (~3.9) and $d$ (~285 nm) were the dielectric constant and thickness of SiO$_2$, respectively).

**Extraction of the contact resistance by using transfer-length method (TLM).** In a two-terminal electronic device, the major resistance contained contact resistance ($R_C$) and channel resistance ($R_{CH}$). Therefore, the total resistance ($R_{TOT}$) normalized by the channel width ($W$) could be described as $R_{TOT} = 2R_C + R_{CH} = 2R_C + R_{SH} \times L_{CH}$, where $R_{SH}$ was the sheet resistance of channel and $L_{CH}$ was the channel length. By measuring the $R_{TOT}$ with various $L_{CH}$, the residual resistance at $L_{CH} = 0$ corresponds to the total contact resistance ($2R_C$) of the device was obtained.

**Extraction of the Schottky barrier.** The contact barrier between 2D semiconductor and metal electrodes could be described by the 2D thermoelectric emission theory:

$$I_{DS} = A_{2D}^* T^{1.5} exp\left(\frac{\Phi_B}{k_B T}\right)\left[1 - exp\left(\frac{-V_{DS}}{k_B T}\right)\right]$$

Where $A_{2D}^* = q(8\pi k_B^3 m^*)^{0.5}/h^2$ was the effective Richardson constant, $q$ was the elementary charge, $k_B$ was the Boltzmann's constant, $T$ was the temperature, and $\Phi_B$ was the Schottky barrier height. This equation could be translated into the Arrhenius equation as $V_{DS} \gg k_B T$, and the barrier height was achieved by extracting the slope of the Arrhenius curve:

$$n(I_{DS}/T^{1.5}) = \frac{-\Phi_B}{k_B T} + C$$

Where $I_{DS}$ was the drain-source current, $C$ was a constant, and $\Phi_B$ was the barrier height derived from the Arrhenius plot. The $\Phi_B$ was then extracted from the flat band gate voltage.

**DFT calculations.** All DFT calculations were performed by using the Vienna *ab* initio simulation package (VASP) with Projector Augmented Wave (PAW) potentials[47,48]. The spin-polarized Generalized Gradient Approximation (GGA) was parametrized by Perdew-Burke-Ernzerhof (PBE)[49], and the cutoff energy for plane wave expansion was set as 450 eV. All structures were optimized by using a force convergence criterion of 0.05 eV/Å and an electronic self-consistency convergence criterion of less than $10^{-5}$ eV. In addition, the empirical dispersion correction using DFT-D3 method was also carried out to describe the possible interaction between $MoS_2$ and $Al_2O_3$ surface. A large lattice length (25 Å) along $z$ axis was used to avoid the interaction between two neighboring images. Based on experimental results, the undoped and Fe-doped $Mo_6S_{14}$ clusters were built on $c$-plane sapphire. Eight Al/O layers were simulated as the sapphire surface with the bottom four layers fixed, top layers of O and Al on the surface were fully relaxed in all of simulations. An Al/O terminated (0001) surface was considered. One row Al/O atoms were deleted to form a stepped surface on $c$-plane sapphire.

**Data availability**

The data that support the findings of this study are available from the corresponding authors upon reasonable request.

**References**

1. Akinwande, D. *et al*. Graphene and two-dimensional materials for silicon technology. *Nature* **573**, 507–518 (2019).
2. Li, M.-Y., Su, S.-K., Philip Wong, H.-S. & Li, L.-J., How 2D semiconductors could extend Moore's law. *Nature* **567**, 169–170 (2019).
3. Liu, Y. *et al*. Promises and prospects of two-dimensional transistors. *Nature* **591**, 43–53 (2021).
4. Wu, F. *et al*. Vertical $MoS_2$ transistors with sub-1-nm gate lengths. *Nature* **603**, 259–264 (2022).
5. Desai, S. B. *et al*. $MoS_2$ transistors with 1-nanometer gate lengths. *Science* **354**, 99–102 (2016).
6. Chhowalla, M., Jena, D. & Zhang, H. Two-dimensional semiconductors for transistors. *Nat. Rev. Mater.* **1**, 16052 (2016).
7. Liu, C. S. *et al*. Two-dimensional materials for next-generation computing technologies. *Nat. Nanotechnol.* **15**, 545–557 (2020).
8. Ahmed, Z. *et al.* Introducing 2D-FETs in device scaling roadmap using DTCO. In *2020 IEEE International Electron Devices Meeting (IEDM)* 22.25.21–22.25.24 (IEEE, 2020).
9. Bubnova, O. 2D materials grow large. *Nat. Nanotechnol.* **16**, 1179 (2021).
10. Dong, J. C., Zhang, L. N., Dai, X. Y. & Ding, F. The epitaxy of 2D materials growth. *Nat. Commun.* **11**, 5862 (2020).

11. Gong, Y. J. *et al*. Spatially controlled doping of two-dimensional $SnS_2$ through intercalation for electronics. *Nat. Nanotechnol.* **13**, 294–299 (2018).

12. Shi, W. *et al*. Reversible writing of high-mobility and high-carrier-density doping patterns in two-dimensional van der Waals heterostructures. *Nat. Electron.* **3**, 99–105 (2020).

13. Liu, Y. *et al*. Approaching the Schottky-Mott limit in van der Waals metal-semiconductor junctions. *Nature* **557**, 696–700 (2018).

14. Wang, Y. *et al*. Van der Waals contacts between three-dimensional metals and two-dimensional semiconductors. *Nature* **568**, 70–74 (2019).

15. Shen, P.-C. *et al*. Ultralow contact resistance between semimetal and monolayer semiconductors. *Nature* **593**, 211–217 (2021).

16. Kwon, G. *et al*. Interaction- and defect-free van der Waals contacts between metals and two-dimensional semiconductors. *Nat. Electron.* **5**, 241–247 (2022).

17. Liu, G. Y. *et al*. Graphene-assisted metal transfer printing for wafer-scale integration of metal electrodes and two-dimensional materials. *Nat. Electron.* **5**, 275–280 (2022).

18. Lee, J.-H. *et al*. Wafer-scale growth of single-crystal monolayer graphene on reusable hydrogen-terminated germanium. *Science* **344**, 286–289 (2014).

19. Wang, M. H. *et al*. Single-crystal, large-area, fold-free monolayer graphene. *Nature* **596**, 519–524 (2021).

20. Li, J. Z. *et al*. Wafer-scale single-crystal monolayer graphene grown on sapphire substrate. *Nat. Mater.* **21**, 740–747 (2022).

21. Lee, J. *et al*. Wafer-scale single-crystal hexagonal boron nitride film *via* self-collimated grain formation. *Science* **362**, 817–821 (2018).

22. Wang, L. *et al*. Epitaxial growth of a 100-square-centimetre single-crystal hexagonal boron nitride monolayer on copper. *Nature* **570**, 91–95 (2019).

23. Chen, T.-A. *et al*. Wafer-scale single-crystal hexagonal boron nitride monolayers on Cu(111). *Nature* **579**, 219–223 (2020).

24. Ma, K. Y. *et al*. Epitaxial single-crystal hexagonal boron nitride multilayers on Ni(111). *Nature* **606**, 88–93 (2022).

25. Xu, X. L. *et al*. Seeded 2D epitaxy of large-area single-crystal films of the van der Waals semiconductor 2H $MoTe_2$. *Science* **372**, 195–200 (2021).

26. Li, T. T. *et al*. Epitaxial growth of wafer-scale molybdenum disulfide semiconductor single crystals on sapphire.

*Nat. Nanotechnol.* **16**, 1201–1207 (2021).

27. Wang, J. H. *et al*. Dual-coupling-guided epitaxial growth of wafer-scale single-crystal WS$_2$ monolayer on vicinal *a*-plane sapphire. *Nat. Nanotechnol.* **17**, 33–38 (2022).

28. Liu, L. *et al*. Uniform nucleation and epitaxy of bilayer molybdenum disulfide on sapphire. *Nature* **605**, 69–75 (2022).

29. Lee, D. *et al*. Remote modulation doping in van der Waals heterostructure transistors. *Nat. Electron.* **4**, 664–670 (2021).

30. Shen, P.-C. *et al*. Healing of donor defect states in monolayer molybdenum disulfide using oxygen-incorporated chemical vapour deposition. *Nat. Electron.* **5**, 28–36 (2022).

31. Meng, W. Q. *et al*. Three-dimensional monolithic micro-LED display driven by atomically thin transistor matrix. *Nat. Nanotechnol.* **16**, 1231–1236 (2021).

32. Lin, H. H. *et al*. Growth of environmentally stable transition metal selenide films. *Nat. Mater.* **18**, 602–607 (2019).

33. Cuccureddu, F. *et al*. Surface morphology of *c*-plane sapphire (α-alumina) produced by high temperature anneal. *Sur. Sci.* **604**, 1294–1299 (2010).

34. Li, B. *et al*. A two-dimensional Fe-doped SnS$_2$ magnetic semiconductor. *Nat. Commun.* **8**, 1958 (2017).

35. Li, H. *et al*. Reducing contact resistance and boosting device performance of monolayer MoS$_2$ by in situ Fe doping. *Adv. Mater.* **34**, 2200885 (2022).

36. Yang, P. F. *et al*. Epitaxial growth of inch-scale single-crystal transition metal dichalcogenides through the patching of unidirectionally orientated ribbons. *Nat. Commun.* **13**, 3238 (2022).

37. Choi, S. H. *et al*. Epitaxial single-crystal growth of transition metal dichalcogenide monolayers *via* the atomic sawtooth Au surface. *Adv. Mater.* **33**, 2006601 (2021).

38. Yang, P. F. *et al*. Epitaxial growth of centimeter-scale single-crystal MoS$_2$ monolayer on Au(111). *ACS Nano* **14**, 5036–5045 (2020).

39. Zhang, L. N., Dong, J. C. & Ding, F. Strategies, status, and challenges in wafer scale single crystalline two-dimensional materials synthesis. *Chem. Rev.* **121**, 6321–6372 (2021).

40. Wan, Y. *et al*. Wafer-scale single-orientation 2D layers by atomic edge-guided epitaxial growth. *Chem. Soc. Rev.* **51**, 803–811 (2022).

41. Liu, Z. *et al*. Strain and structure heterogeneity in MoS$_2$ atomic layers grown by chemical vapour deposition. *Nat. Commun.* **5**, 5246 (2014).

42. Zeng, Y. *et al*. Highly enhanced photoluminescence of monolayer MoS$_2$ with self-assembled Au nanoparticle arrays. *Adv. Mater.* **4**, 1700739 (2017).

43. Lin, Z. Y. *et al*. Solution-processable 2D semiconductors for high-performance large-area electronics. *Nature* **562**, 254–258 (2018).

44. Li, N. *et al*. Large-scale flexible and transparent electronics based on monolayer molybdenum disulfide field-effect transistors. *Nat. Electron.* **3**, 711–717 (2020).

45. Sebastian, A., Pendurthi, R., Choudhury, T. H., Redwing, J. M. & Das, S. Benchmarking monolayer MoS$_2$ and WS$_2$ field-effect transistors. *Nat. Commun.* **12**, 693 (2021).

46. Chowdhury, T. et al. Substrate-directed synthesis of MoS$_2$ nanocrystals with tunable dimensionality and optical properties. *Nat. Nanotechnol.* **15**, 29–34 (2020).

47. Kresse, G. & Furthmuller, J. Efficiency of *ab*-initio total energy calculations for metals and semiconductors using a plane-wave basis set. *Comput. Mater. Sci.* **6**, 15–50 (1996).

48. Kresse, G. & Furthmuller, J. Efficient iterative schemes for *ab* initio total-energy calculations using a plane-wave basis set. *Phys. Rev. B* **54**, 11169–11186 (1996).

49. Perdew, J. P., Burke, K. & Ernzerhof, M. Generalized gradient approximation made simple. *Phys. Rev. Lett.* **77**, 3865–3868 (1996).

**Acknowledgements**

This work is supported by the National Key R&D Program of China (grant nos. 2018YFA0703700 and 2021YFA1200800), the National Natural Science Foundation of China (grant nos. 91964203 and 92164103), the Beijing National Laboratory for Molecular Sciences (grant no. BNLMS202001), and the Fundamental Research Funds for the Central Universities (grant no. 2042021kf0029). The authors would like to acknowledge the Center for Electron Microscopy at Wuhan University for their substantial supports to JEM-F200, JEM-NEOARM, and JEM-ARM200F.

**Author contributions**

J.S. and J.H. conceived and supervised the project. H.L., J.Y. and X.L. performed CVD growth with M.C., W.F., R.D., Y.W., L.S., X.W. and Y.Z.'s assistance. H.L. and L.L. constructed the device and electrical transport measurements. H.L., J.Y., X.L., M.C., W.F., R.D., Y.W., L.S. and X.W. carried out OM, XPS, Raman, AFM, TEM characterizations and data analysis. J.S. and J.H. co-wrote the manuscript, with input from the other authors. All authors contributed to discussions.

## Competing interests

The authors declare no competing interests.

## Additional information

**Supplementary information** The online version contains supplementary material available at XXX.

**Correspondence** and requests for materials should be addressed to J. S. or J. H.

**Reprints and permission information** is available at XXX.